\documentclass{icrc29}
\usepackage{graphicx,amssymb,amsmath,times}
\begin{document}
\title[The MINOS Detectors]{The MINOS Detectors}
\author[A. Habig, E.W. Grashorn et al.] {A. Habig$^a$, E.W. Grashorn$^a$,
  for the MINOS Collaboration$^b$\\
        (a) Univ. of Minnesota Duluth Physics Dept., 10 University Dr.,
        Duluth, MN 55812, USA\\
        (b) http://www-numi.fnal.gov/collab/collab.ps
        }
\presenter{Presenter: A. Habig (ahabig@umn.edu), \  
usa-habig-A-abs1-he24-poster}

\maketitle

\begin{abstract}

The Main Injector Neutrino Oscillation Search (MINOS) experiment's
primary goal is the precision measurement of the neutrino oscillation
parameters in the atmospheric neutrino sector.  This long-baseline
experiment uses Fermilab's NuMI beam, measured with a Near Detector at
Fermilab, and again 735~km later using a Far Detector in the Soudan Mine
Underground Lab in northern Minnesota.  The detectors are magnetized
iron/scintillator calorimeters.  The Far Detector has been operational
for cosmic ray and atmospheric neutrino data from July of 2003, the Near
Detector from September 2004, and the NuMI beam started in early
2005.  This poster presents details of the two detectors.

\end{abstract}

\section{Introduction} The MINOS experiment uses two similar detectors
to measure the properties of the NuMI neutrino beam over a long
baseline, in order to precisely measure the neutrino flavor oscillations
seen in atmospheric neutrinos~\cite{SK-full}.  The beam is created using
protons from Fermilab's Main Injector and is characterized by the Near
Detector, located 1~km from the target and 90~m under the surface at
Fermilab.  The beam then travels northwest 735~km to the Far Detector,
located 700~m (2070~mwe) deep in the Soudan Mine Underground Lab in
northern Minnesota.

The two detectors are designed to be as similar as possible to minimize
the systematic errors resulting from a comparison of the observed
neutrino spectra in the two detectors.  Both are constructed of
vertically hung planes, each plane having a layer of steel (to act as
both a neutrino target and a calorimeter) and a layer of plastic
scintillator to detect passing charged particles.  The steel is also
magnetized to allow charge discrimination and momentum determination.
The planes are oriented perpendicular to the path of the beam.

\section{The Near and Far Detectors}

\begin{figure}[h]
  \begin{center}
    \begin{minipage}[c]{0.4\textwidth}
      \includegraphics[width=\textwidth]{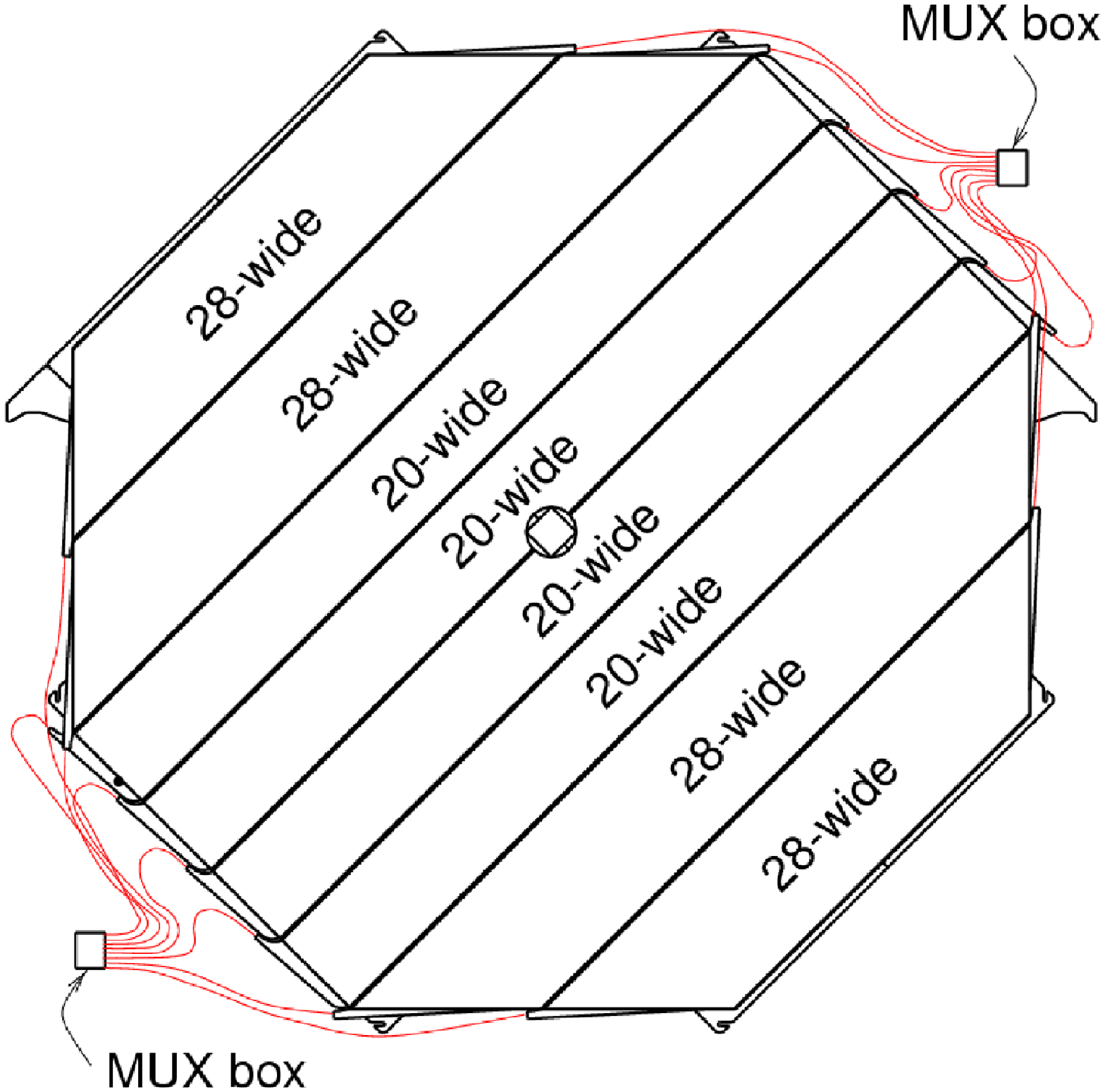}
    \end{minipage}
    \begin{minipage}[c]{0.4\textwidth}
      \includegraphics[width=\textwidth]{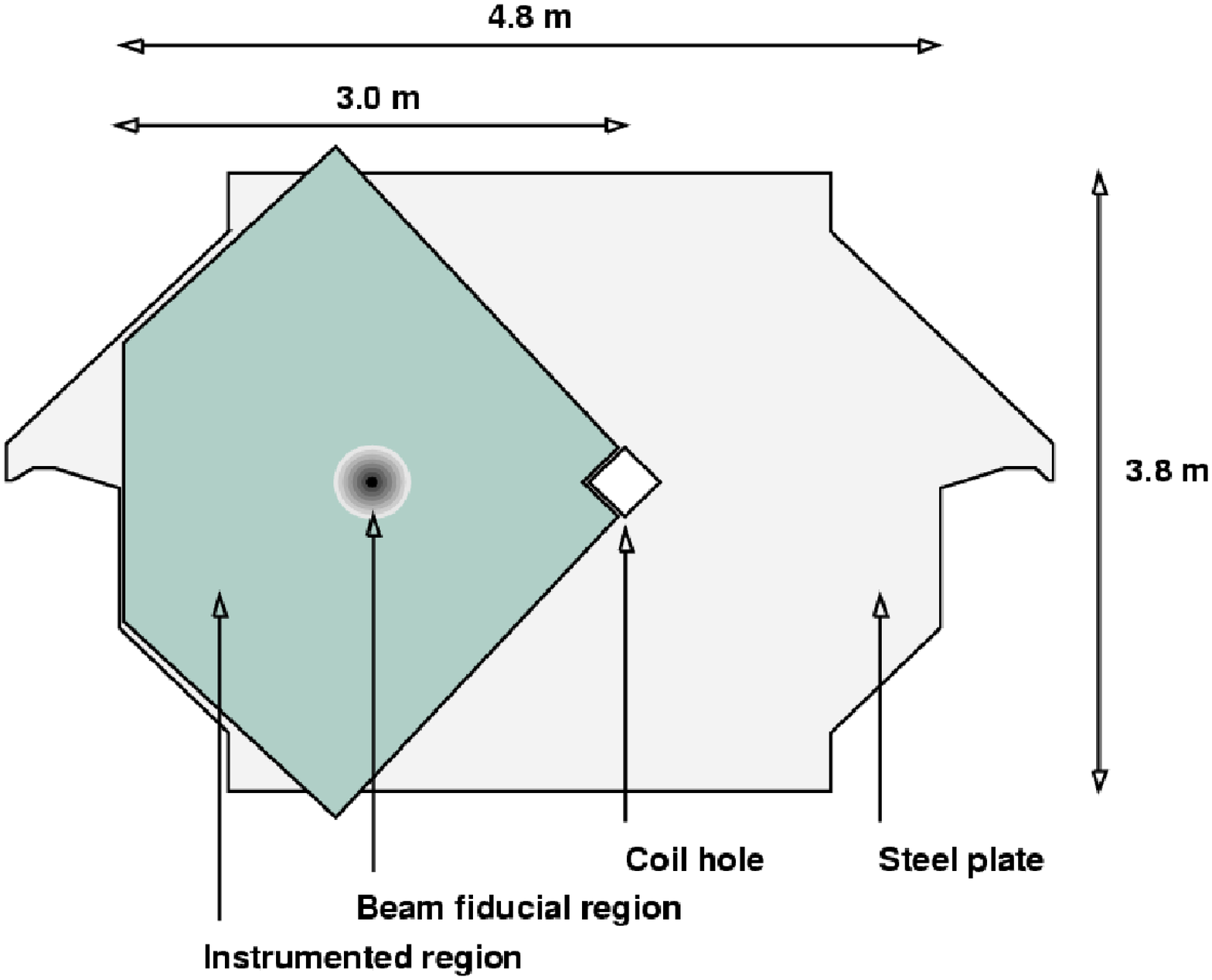}
    \end{minipage}

    \caption{\label{fig:planes} 
      A diagram of one plane of the the MINOS Far~(l) and Near~(r)
      Detectors.  A Far Detector plane is 8~m in diameter, a Near
      Detector plane is 3.8~m tall and 4.8~m wide.  The ``ears'' on the
      sides support the weight of the hanging plane, and an
      electromagnetic coil passes through the hole in the center of each
      plane.  The Far planes are instrumented across the whole area (in
      this example the modules of scintillator are seen running from
      lower left to upper right), in the Near some are almost fully
      instrumented others only on the side hit by the beam (the
      diamond-shaped region on the left).  The beam center is shown by the
      circle in the center-left of the Near plane.
    }
  \end{center}
\end{figure}

The Far Detector is composed of 486 planes and has a total mass of 5400
tons.  It is octagonal in cross section (Fig.~\ref{fig:planes})
with a diameter of 8~m and is divided into two halves or ``Supermodules'',
each 15~m in depth and separated by a 1.5~m air gap.  In addition, a
veto shield composed of two layers of scintillator (but no steel) is
placed over the top and sides of the Far Detector, to better identify
incoming cosmic ray muons that might otherwise enter down the cracks
between planes and suddenly appear inside the detector's fiducial
volume.

The Near Detector geometry has a 3.8~m$\times$4.8~m squashed octagon
cross section (Fig.~\ref{fig:planes}).  As the beam is centered on the
left of the detector and a fiducial region of radius 1~m defined about
this point, not all the remaining area is instrumented.  There
are 282 planes (153 instrumented) for a total mass of 980 tons, spanning
a length of 16.6~m.  The detector is divided into two regions.  The 120
planes in the upstream ``forward'' region are all instrumented.  Every
fifth plane has scintillator over most of the plane's area, with the
scintillator on the rest only covering the left side where the beam
crosses.  This region is used to get a detailed picture of the neutrino
interactions occurring in its volume.  The 162 planes downstream in the
``spectrometer'' section are only instrumented every fifth plane (albeit
with full scintillator coverage).  This region's purpose is to track the
penetrating muons generated by interactions upstream for a momentum
determination of these long tracks.

\subsection{Steel}

The steel planes are 2.54~cm thick AISI 1006 low carbon steel (carbon
content between 0.04\% and 0.06\%), flat to within $\pm$0.76~mm.  Each
detector is magnetized for charge identification and to determine a
particle's momentum by measuring the resulting curvature of its track.
The current runs through a coil passing through a 25~cm hole in
center of the detector.  The Far Detector's coil provides 15~kA-turns,
the Near coil 40~kA-turns, resulting in a toroidal magnetic field in the
steel varying between 1--2~T as a function of radius.  Knowing how the
magnetic field varies in the detector is important to the accuracy of
the results obtained from the track curvature.  This is accomplished by
detailed knowledge of the steel chemistry and careful modeling,
cross-checked by measurements of the induced current in induction loops
built around each plane of steel.

\subsection{Scintillator}

The scintillator is composed of $4\times1$~cm polystyrene strips
extruded to length (up to 8~m in the Far Detector)~\cite{scint}.  The
plastic is doped with 1\% PPO plus 0.03\% POPOP fluors and co-extruded
with a white TiO$_2$ layer to contain scintillation light.  A groove is
cut in this surface down one wide side of the strip and a 1.2~mm Kuraray
wavelength shifting (``WLS'') fiber (containing Y-11 at 175ppm) glued
into the groove to collect the light.  Clear fiber carries the light
from the ends of the strips to the phototubes.  Also at the strip ends
another fiber brings in light from a UV LED.  This ``Light Injection''
system allows the calibration and verification of the complete system by
injecting known quantities of light into specific parts of the
detector~\cite{LI}.
Strips are laid side-by-side and bundled into ``modules'', wrapped
with an aluminum skin for light-tightness and ease of handling.  Both
ends of the Far Detector strips are read out but only one end at the
Near.  The strips are oriented at $45^\circ$ from vertical.  Alternating
planes have strips orthogonal to the last to allow stereo readout.

\subsection{Electronics}

The light gathered by the clear fibers is carried to ``Front End''
electronics racks, where it enters a dark box to be optically coupled to
a photomultiplier tube (``PMT'').  At the Far Detector, light from eight
different strip ends is combined onto one pixel of a Hammamatsu
R6000-M16 (Far)~\cite{M16-PMTs} multi-pixel PMT.  M-64 PMTs are used at
the Near~\cite{M64-PMTs}.  There is no multiplexing in the forward
region of the Near Detector due to its small size and higher rates,
although the spectrometer region is four-way multiplexed.

The phototube signals are digitized by Viking ``VA'' chips at the Far
Detector~\cite{fardet-fe} and FNAL ``QIE'' chips to meet the faster
demands of the high event rate at the Near.  In both cases, circuitry is
installed to inject known amounts of charge into the digitization
circuits for calibration of the digitization.  Data are gathered into
custom mid-level VME cards for readout by the data
acquisition~\cite{DAQ}.  The unexpectedly high singles rate from light
generated in the WLS~\cite{WLS} is mitigated by the imposition of a 2/36
coincidence cut in these mid-level electronics at the Far Detector, but
the faster electronics at the Near Detector has no problem with the
rates.  The data acquisition reads charge and time information (called
``digits'') for each phototube signal, combines such data from all parts
of the detector, and arranges them into time-ordered ``snarls'' of data
that correspond to physical events.  At the time of a beam spill all
data is kept.  The Near Detector uses the accelerator's RF signal to
gate the low-level electronics, and the far gets GPS beam spill
timestamps via the internet and applies this to buffered data in the
DAQ's trigger processing farm.  Absent a beam spill, low level singles
noise is further suppressed by the application of a software trigger
requiring either 4/5 contiguous planes hit (for tracks and large
showers) or 6 hits and 1500 ADC counts in any 4 plane window (for small
showers).  The resulting data is written to disk in the ``Root''
format~\cite{root} for later analysis.

\subsection{Operation and Performance}

The Far Detector has been fully operational since July 2003, while the
Near planes were fully commissioned in August 2004.  During the Near
Detector installation, as individual planes were commissioned during
installation they were added to the data acquisition piecemeal, so a
larger set of data exists for partial detector configurations.  Cosmic
ray muons are seen in both detectors, at rates of 0.5~Hz (Far) and
270~Hz (Near).  These muons are used to calibrate the detector and
validate the data acquisition and reconstruction, as well as studied in
their own right.  The seasonal variations in the cosmic ray muon rate
has been seen, as has the shadow cast by the Moon in the primary cosmic
ray sky.  The Far Detector's veto shield is 97\% efficient at rejecting
cosmic rays, allowing the detection of atmospheric neutrinos.

\begin{figure}[h]
  \begin{center}
    \begin{minipage}[c]{0.4\textwidth}
      \includegraphics[width=\textwidth]{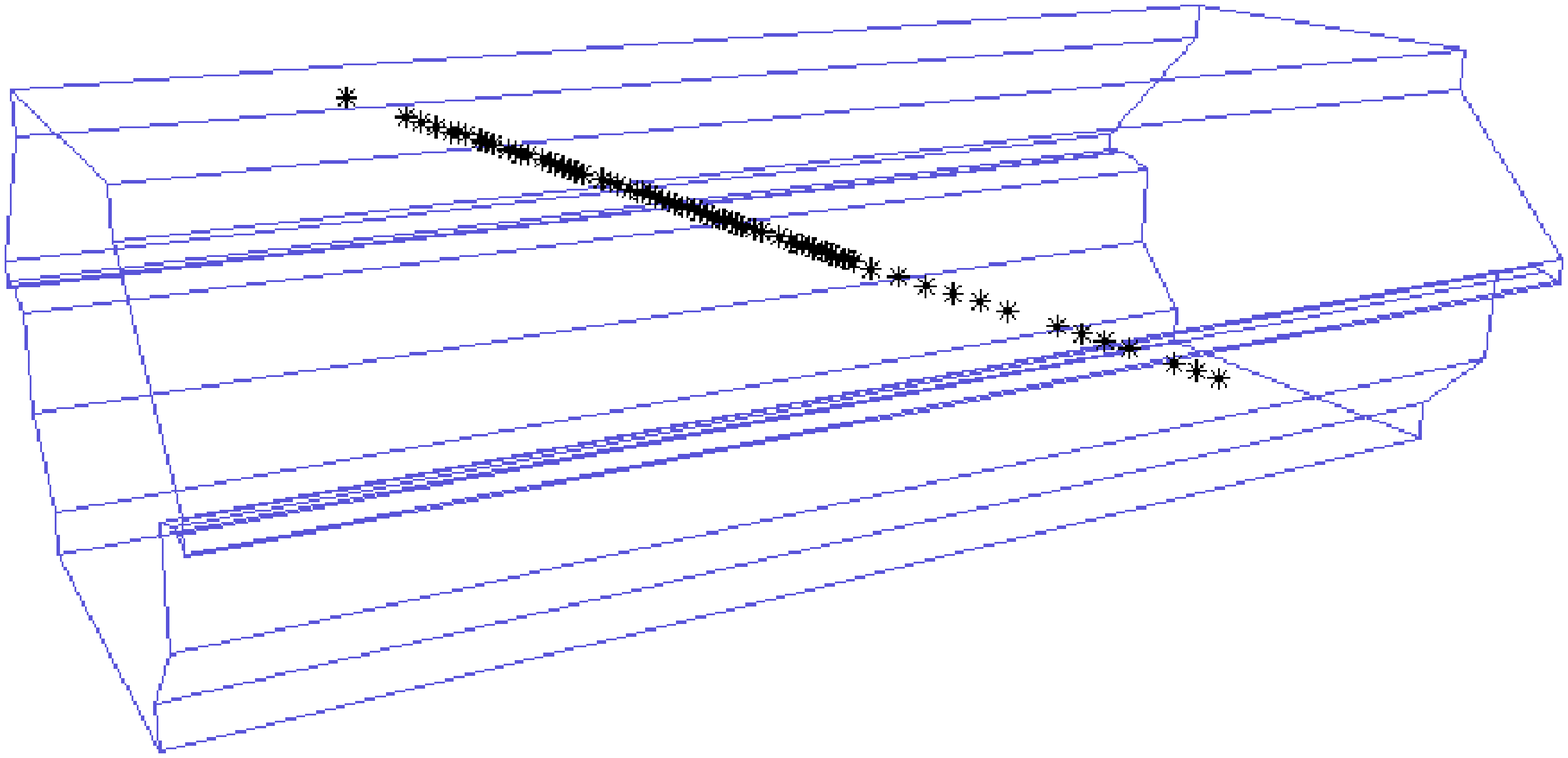}
    \end{minipage}
    \begin{minipage}[c]{0.4\textwidth}
      \includegraphics[width=\textwidth]{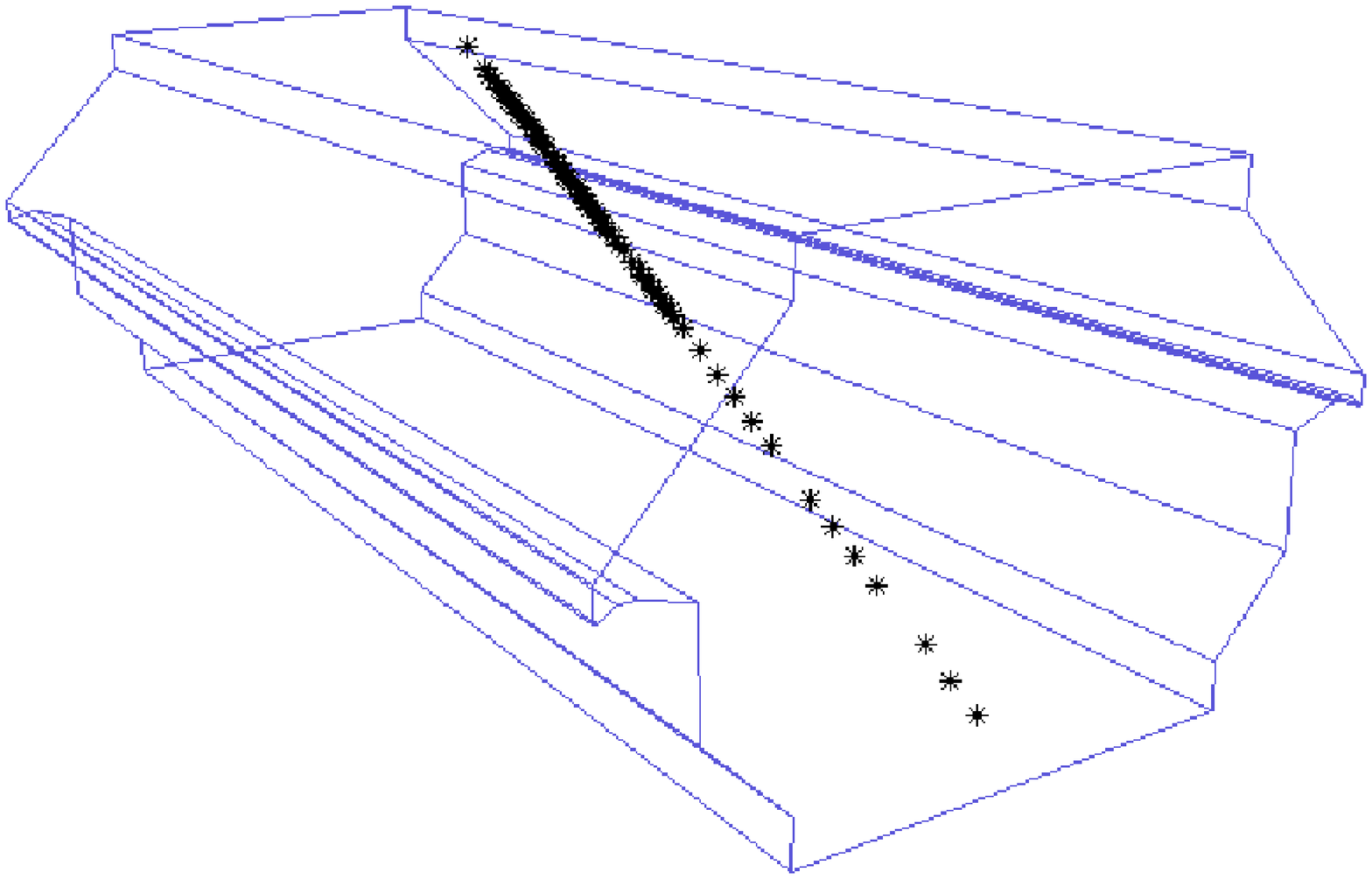}
    \end{minipage}

    \caption{\label{fig:evd} 
      A cosmic-ray muon seen crossing the MINOS Far Detector, in two
      rotated 3D views.
    }
  \end{center}
\end{figure}

A particle interaction in the detector can be localized to 5.94~cm in
the longitudinal direction and 4.1~cm in the transverse direction,
allowing muon tracks to be resolved to $0.25^\circ$.  The energy
resolution for showers is 55\%/$\sqrt{E}$ for hadrons and
23\%/$\sqrt{E}$ for electrons.  A muon typically yields nine
photoelectrons per plane with a timing resolution of 2.4~ns.  An example
of a cosmic-ray muon track in the Far Detector can be seen in
Fig.~\ref{fig:evd}.  The Far Detector averages over 95\% live time for
cosmic rays, and well above 99\% for beam data.

The Near Detector sees around 20 neutrino interactions during a typical
$8.67\mu$s long beam spill, during which data is digitized
continuously for 18~ms.  The resulting data from individual neutrino
interactions are ``sliced'' into discrete events offline, with a
granularity of 19~ns.  The Far Detector sees far less than one event per
spill on average.  Neutrino interactions have been observed in the Near
Detector since the first beam tests in January 2005, and in the Far
Detector since the beam intensity was increased to operational levels in
March 2005, and continue to be collected.

\section{Conclusions}

The Near and Far Detectors of the MINOS long-baseline neutrino
oscillation experiment are complete and operational.  The two detectors
are similar layered iron/scintillator calorimeters to minimize
systematic differences.  The plastic scintillator is extruded in strips
and read out via optical fibers carrying the scintillation light to
photomultiplier tubes. Cosmic ray data have been taken in both
detectors, starting July 2003 at the Far Detector and August 2004 at the
Near.  Atmospheric neutrino events have been observed at the Far
Detector at a rate of a couple per week~\cite{icrc-atmnu}.  Beam
neutrinos have been observed in both detectors, as the beam has been in
an operational mode since March of 2005.  The MINOS detectors are
performing as expected.

\section{Acknowledgments}

This work was supported by the U.S. Department of Energy, the U.K.
Particle Physics and Astronomy Research Council, and the State and
University of Minnesota.  We gratefully acknowledge the Minnesota
Department of Natural Resources for allowing us to use the facilities of
the Soudan Underground Mine State Park.  This presentation was directly
supported by NSF RUI grant \#0354848.

\end{document}